\documentclass[twocolumn,preprintnumbers,amsmath,amssymb,floatfix,superscriptaddress,nofootinbib]{revtex4}
\usepackage{amsmath,hyperref,amssymb}
\usepackage{amsmath}
\usepackage{graphicx}% Include figure files
\usepackage{dcolumn}% Align table columns on decimal point
\usepackage{bm}% bold math
\usepackage{verbatim}
\usepackage{tabularx}
\usepackage{slashed}
\usepackage{cancel}
\usepackage{hyperref}
\usepackage{comment}
\usepackage{color}

\def\be{\begin{equation}}
\def\ee{\end{equation}}
\def\bea{\begin{eqnarray}}
\def\eea{\end{eqnarray}}
\def\ba#1\ea{\begin{align}#1\end{align}}
\def\bg#1\eg{\begin{gather}#1\end{gather}}
\def\bm#1\em{\begin{multline}#1\end{multline}}
\def\bmd#1\emd{\begin{multlined}#1\end{multlined}}

\renewcommand{\t}{\tilde}
\newcommand{\tr}{\text{tr}}
\newcommand{\sgn}{{\rm{sgn}}}

\renewcommand{\(}{\left(}
\renewcommand{\)}{\right)}
\renewcommand{\[}{\left[}
\renewcommand{\]}{\right]}
\renewcommand{\Re}{\textrm{Re}}

\begin{document}

\title{Superconductivity vs quantum criticality: effects of thermal fluctuations}
%\addYW{Superconductivity vs non-Fermi liquid} at finite temperature: effects of thermal fluctuations
\author{Huajia Wang}
\thanks{These authors have contributed equally to this work.}
\affiliation{Department of Physics, University of Illinois, Urbana IL, USA}
\author{Yuxuan Wang}
\thanks{These authors have contributed equally to this work.}
\affiliation{Department of Physics, University of Illinois, Urbana IL, USA}
\author{Gonzalo Torroba}
\thanks{torrobag@gmail.com}
\affiliation{Centro At\'omico Bariloche and CONICET, Bariloche, Rio Negro R8402AGP,
Argentina}

\date{\today}
\begin{abstract}
We study the interplay between superconductivity and non-Fermi liquid behavior of a Fermi surface coupled to a massless $SU(N)$ matrix boson near the quantum critical point. The presence of thermal infrared singularities in both the fermionic self-energy and the gap equation invalidates the Eliashberg approximation, and makes the quantum-critical pairing problem qualitatively different from that at zero temperature. Taking the large $N$ limit, we solve the gap equation beyond the Eliashberg approximation, and obtain the superconducting temperature $T_c$ as a function of $N$. Our results show an anomalous scaling between the zero-temperature gap and $T_c$. For $N$ greater than a critical value, we find that $T_c$ vanishes with a Berezinskii-Kosterlitz-Thouless scaling behavior, and the system retains non-Fermi liquid behavior down to zero temperature. This confirms and extends previous renormalization-group analyses done at $T=0$, and provides a controlled example of a naked quantum critical point. We discuss the crucial role of thermal fluctuations in relating our results with earlier work where superconductivity always develops due to the special role of the first Matsubara frequency.

\end{abstract}

\maketitle

%%%%%%%%%%%%%%%%%%%%%%%%%%%%%%%%%%%%%%
%%%%%%%%%%%%%%%%%%%%%%%%%%%%%%%%%%%%%%
\section{Introduction}\label{sec:intro}

One of the key open problems in modern condensed matter physics is the relation between superconductivity (SC) and non-Fermi liquid (NFL) behavior. The interplay between these phenomena may be partly responsible for the existence of high temperature superconductivity, and could also describe some striking transport properties of strongly correlated materials. This is strongly supported by recent experimental~\cite{Shibauchi2013,analytis, boebinger1, boebinger2, Zhang:2016ofh} and numerical~\cite{yoni1,yoni2,yoni3,meng1,meng2} measurements of NFL signatures, and the astonishing evidence of quantum critical points hidden behind superconducting domes~\cite{Stewart2001, Broun2008, Gegenwart2008, taillefer-review, ramshaw,taillefer-new}.

Systems near quantum criticality are expected to have a field theory description in terms of a Fermi surface interacting with additional soft bosons, which typically arise e.g. from order parameter fluctuations. The exchange of virtual bosons has a dual effect: it enhances the fermionic pairing interaction, and it also gives rise to NFL behavior (a fermionic anomalous dimension) that tends to make SC irrelevant. If the enhancement of pairing dominates, it will lead to a parametric increase in the SC critical temperature $T_c$~\cite{Son:1998uk}. However, if the anomalous dimension effect can be made large enough, it may produce a NFL-like superconductor (with pairing formed by incoherent fermions), or even the possibility of a naked quantum critical point (i.e., a quantum critical point not preempted by a SC order) down to zero temperature.

While very appealing, this idea faces a basic problem: in most field-theoretical models, NFL effects do not set in before the SC instability. Instead, a common outcome in models that are under perturbative control is that $\Delta \gg \Lambda_{\rm NFL}$, where $\Delta$ is the scale of the SC gap, and $\Lambda_{\rm NFL}$ is the crossover scale at which NFL effects become important. A paradigmatic example of this situation is the large-$N$-component fermion limit~\cite{Polchinski1994, Altshuler1994}. See Ref.~\cite{Metlitski} for an analysis of some of the possibilities. 
On the other hand, a recent renormalization group analysis at $T=0$ shows~\cite{Raghu:2015sna} (see also~\cite{acf} for an earlier related work) that superconductivity can be completely suppressed  compared to the NFL behavior in the class of models with an $N\times N$ matrix boson studied by~\cite{Fitzpatrickone, FKKRtwo, Torroba:2014gqa}.
 However, at \emph{finite temperature}, the recent work~\cite{2016PhRvL.117o7001W} emphasized unique low-energy effects beyond previous renormalization group (RG) approaches, and argued that as a result,
pairing always wins over NFL behavior. This  physics at {finite} temperatures has not been addressed before. For these reasons, it becomes necessary to reconsider
quantum critical pairing at finite temperature within the framework of field theory.

In this work we study the interplay between pairing and quantum criticality at finite temperature. For this, we focus on the aforementioned class of models developed in~\cite{Fitzpatrickone, FKKRtwo, Torroba:2014gqa, Raghu:2015sna}, which feature an $SU(N)$ global symmetry (a generalization of the spin symmetry group). The fermion has $N$ components but, unlike vector models \cite{Polchinski1994, Altshuler1994} where the boson is a singlet, here the boson is an $N \times N$ matrix (adjoint representation). This is a central point in order to allow for NFL effects to dominate.
At zero temperature, increasing $N$ tends to make NFL effects stronger, and it is possible to obtain critical pairing interactions~\cite{Raghu:2015sna, Wang:2016hir}. Building on the recent normal-state analysis~\cite{Wang:2017kab}, we develop analytic and numerical methods to explore the SC and quantum critical regimes, and obtain the phase diagram as a function of $(T, N)$. 

For conventional BCS superconductivity, the pairing problems at finite $T$ and $T=0$ are qualitatively the same, with $T_c$ of the order of the pairing gap $\Delta$ at $T=0$. However, in our case the situation is fundamentally different. At finite $T$, the frequency integral in the gap equation is replaced by a Matsubara sum. Since the pairing interaction is mediated by a massless boson, the standard Eliashberg approximation (briefly reviewed in Sec.\ \ref{sec:beyondeliash}) leads to an infrared divergence from the term with bosonic frequency $\Omega_m=0$ (i.e., the thermal piece). This issue actually afflicts many quantum-critical pairing problems, but has not attracted much attention for two reasons. First, it is known that for $s$-wave pairing with fermion flavor $N=1$, this divergence is canceled by a similar divergence in the fermionic self-energy~\cite{acn, 2016PhRvL.117o7001W}, via an analog of Anderson's theorem for impurities~\cite{anderson,acn}.
Second, even for cases without exact cancellation, the issue is generally ignored as the usual approach is to simply replace the Matsubara sum by an integral, $T\sum \to \int d\omega/2\pi$. Indeed, the \emph{integral} of the divergence at $\Omega_m=0$ is convergent and there appears to be no IR problem. We will show, however, that for our case the infrared singularities invalidate the Eliashberg approximation and the replacement of the sum by an integral.
One of our key results is to go beyond the Eliashberg approximation at $\Omega_m=0$, showing how the apparent IR divergences are resolved at large $N$. We then solve the linear gap equation to obtain the transition temperature $T_c$. For the cases where $T_c\neq 0$, we find an anomalous scaling relation between the zero-temperature SC gap and $T_c$, 
\be
\frac{\Delta (T=0)}{T_c} \sim T_c^{-{2\epsilon}/{3}}\gg 1\,,
\ee 
where $\epsilon$ is the difference from the upper critical dimension $d=3-\epsilon$. This is distinct from the standard BCS result that $\Delta/T_c\approx 1.76 = O(1)$. Such an anomalous relation has its roots in the qualitative difference between the pairing problems at finite and zero $T$.

Our finite $T$ analysis shows that
 for low $N$ SC is parametrically enhanced by soft boson exchange, in agreement with~\cite{Son:1998uk}. However, as $N$ is increased the NFL effects tend to destroy the SC order. Above a critical value $N_c=12/\epsilon$, $T_c \to 0$, SC is completely extinguished via an infinite order transition (at which all derivatives of observables are smooth), and a naked quantum critical region ensues. This extends the findings of~\cite{Raghu:2015sna, Wang:2016hir} to finite temperature. We will see that the nonperturbative resolution of IR divergences plays a central role here.

The result the SC can be avoided near the quantum critical region is in apparent contradiction with~\cite{2016PhRvL.117o7001W}. The key ingredient there is the special role of first Matsubara frequency ($\omega_m=\pm \pi T$) in calculations of superconductivity at which the fermionic self-energy vanishes. The resolution of the contradiction lies in the fact that in our model, as shown recently in~\cite{Wang:2017kab}, thermal fluctuations dominate the fermionic self-energy at lowest frequencies (for any $N\neq 1$), and this eliminates the special role of the first Matsubara frequency.  To clarify the connection with~\cite{2016PhRvL.117o7001W}, we show how the first-Matsubara effect re-emerges if the thermal contribution to the self-energy is reduced. While at this stage it is not clear how to realize this limit in a controlled fashion, if this does happen we will find that the interplay between first Matsubara physics and NFL thermal effects gives rise to an intriguing reentrance behavior, whereby the system enters and exits the SC phase to return to the NFL normal state at the lowest temperatures.

Our results provide a theoretical proof to the effect that quantum criticality can dominate over pairing instabilities down to zero temperature.  We also emphasize the crucial role of the large $N$ limit for NFLs; as has been the case in other areas of physics (such as quantum chromodynamics), it is likely that large $N$ will also capture qualitative aspects of the physics with a small number of degrees of freedom. Finally,
we hope that our proof of principle stimulates the search for related field theory mechanisms, and the construction of phenomenologically realistic models.

The rest of this paper is structured as follows. In Sec.\ \ref{sec:qc} we review the framework of our model and the treatment of the normal state fermionic self-energy. In particular, due to the breakdown of the Eliashberg approximation at frequency transfer $\Omega_m=\omega_m-\omega_m'=0$, the self-energy contains a new thermal contribution. In Sec.\ \ref{sec:sc}, we study the the pairing problem. As in the normal state analysis, the thermal term with $\Omega_m=0$ requires a special treatment beyond the Eliashberg approximation. We obtain the correct form of the finite-$T$ gap equation and solve it both numerically and analytically. Our finite-$T$ analysis confirms and extends previous findings of metallic NFL behavior at $T=0$. In Sec.\ \ref{sec:mats} we investigate the role of the first Matsubara frequency within our large-$N$ model. We further show that by reducing the thermal contribution in the self-energy, the special role of the first Matsubara frequency emerges and its interplay between the thermal effects discussed above can lead to an interesting re-entrant superconducting behavior. Finally, in Sec.\ \ref{sec:final} we summarize the phase diagram that obtains from our analysis, and discuss future directions. Some explicit calculations are presented in the Appendix.

%%%%%%%%%%%%%%%%%%%%%%%%%%%%%%%%%%%%%%
%%%%%%%%%%%%%%%%%%%%%%%%%%%%%%%%%%%%%%
\section{Quantum critical dynamics in the normal state}\label{sec:qc}

We begin by reviewing the class of non-Fermi liquids that we will study, and its main properties near quantum criticality. We consider a Fermi surface of fermions $\psi$ interacting with a nearly massless boson $\phi$ (e.g. an order parameter fluctuation). The system is generalized to introduce an $SU(N)$ global symmetry  under which $\psi$ is a fundamental and $\phi$ transforms in the adjoint. We consider $N \gg 1$ as a formal limit for solving exactly the quantum dynamics of the model. Furthermore, we perform an $\epsilon$-expansion for $d=3-\epsilon$ spatial dimensions. The euclidean Lagrangian is
\bea\label{eq:L1}
L&=&\frac{1}{2}\tr \left((\partial_\tau \phi)^2+ (\vec 
\nabla \phi)^2 \right)+ \psi_i^\dag \left(\partial_\tau+ \varepsilon(i\vec \nabla)-\mu_F
\right)\psi^i \nonumber\\
&+& \frac{g}{\sqrt{N}} \phi^i_j \psi^\dag_i \psi^j- \lambda_{BCS}\, \psi^\dag_i\psi^j\psi^\dag_j\psi^i\,.
\eea
For simplicity, the dispersion relation $\varepsilon(\vec k)$ is taken to be spherically symmetric; $k_F$ will denote the Fermi momentum, and $v=\varepsilon'(k_F)$ the Fermi velocity.
The third term is the boson-fermion scattering, with $g$ fixed at large $N$. The 4-fermion interaction is evaluated on antipodal points of the Fermi surface (BCS kinematics), and is responsible for the superconducting instability. For simplicity, we will set $\lambda_{BCS}=0$ at the UV cutoff so that, below that scale, pairing is due to boson exchange alone, and is of order $g^2/N$.

\begin{figure}[h]
\begin{center}
\includegraphics[width=\columnwidth]{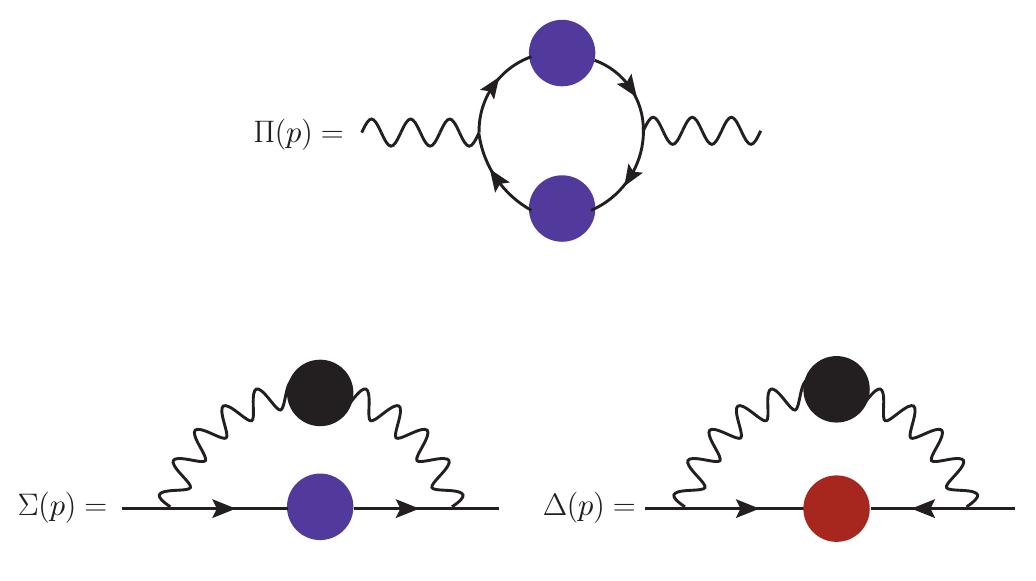} 
\caption{Schwinger-Dyson equations for the full boson propagator (black), and fermionic propagators $\langle \psi \psi^\dag\rangle$ (blue), $\langle \psi \psi\rangle$ (red). The contribution from fermion pairing $\langle \psi \psi\rangle$ to $\Pi$, not included, vanishes at $T_c$. }
\label{fig:SD}
\end{center}
\end{figure}

Large $N$ simplifies considerably the task of calculating quantum correlation functions, because corrections to the Yukawa coupling vertex are suppressed by $O(1/N)$. In this case, the boson self-energy $\Pi$, fermion self-energy $\Sigma$, and fermion SC gap $\Delta$ obey a closed-set of Schwinger-Dyson equations, shown diagrammatically in Fig.~\ref{fig:SD}. The other key property of the large $N$ limit is that the contribution to the self-energy $\Sigma$ appears at leading order in $N$, while for the gap the leading contribution comes from order $1/N$~\cite{Fitzpatrick:2014efa}. (This can also be seen explicitly in terms of the double-line notation commonly adopted in large $N$ gauge theories~\cite{tHooft:1973alw}: the diagrams for $\Sigma$ are planar, while those for $\Delta$ are non-planar.) A competing effect is that tree-level boson exchange gives rise to a non-local 4-Fermi interaction that tends to enhance $\Delta$ exponentially, as in~\cite{Son:1998uk}. Combining both effects we expect that, as $N$ is increased, NFL behavior starts to compete with, and may eventually overcome, the pairing instability.

Our task is to solve these equations at finite temperature. We start from the high-temperature normal state, and search for $T_c$ at which SC develops. The normal-state bosonic and fermionic Green's functions are denoted by
\bea\label{eq:fullG1}
D(\Omega_n, p) &=& \frac{1}{\Omega_n^2 + p^2 +\Pi(\Omega_n, p)}\nonumber\\
G(\omega_n, p) &=& -\frac{1}{i \omega_n +i \Sigma(\omega_n, p)-\varepsilon(p)+\mu_F}\,,
\eea
with Matsubara frequencies $\Omega_n= 2n \pi T$, $\omega_n=(2n+1)\pi T$. As $T \to T_c$ from above,
it is sufficient to linearize the Schwinger-Dyson equations around $\Delta=0$, which obtains
\begin{widetext}
\bea\label{eq:SD}
\Pi(\Omega_n, q) &=& \frac{g^2}{N}\, T \sum_m \int\frac{d^d p}{(2\pi)^d}\,G(\omega_m-\Omega_n, p-q)\,G(\omega_m, p) \nonumber\\
i \Sigma( \omega_n ,q)&=& g^2 T\sum_m \int\frac{d^d p}{(2\pi)^d}\, D(\omega_m-\omega_n, p-q) G(\omega_m, p) \\
Z(\omega_n,q)\Delta(\omega_n,q)&=& \frac{g^2}{N}\, T \sum_m \int\frac{d^d p}{(2\pi)^d}\,D(\omega_m-\omega_n, p-q)   |G(\omega_m,p)|^2 Z(\omega_m,p)\Delta(\omega_m,p)\,. \nonumber
\eea
\end{widetext}
Note that $\Delta(\omega_n,p)$ is the physical gap function, i.e. it enters the Lagrangian of the SC phase as 
\be
L \supset Z (-i \omega \psi_p^\dag \psi_p + \Delta\psi_p \psi_{-p} + \text{c.c.})
\ee
with 
\be
Z(\omega,p) \equiv 1+ \frac{\Sigma(\omega,p)}{\omega}
\ee 
the wavefunction renormalization.

In the linearized approximation, $\Pi$ and $\Sigma$ are independent of the gap, and coincide with the normal state self-energies, which has been analyzed by two of us in~\cite{Wang:2017kab}. See also~\cite{LeBellac:1996kr, Blaizot:1996az, Abanov2003} for previous related work. We will now summarize the results there, and in the following sections we will focus on the gap equation.

The low frequency boson self-energy turns out to be one-loop exact at finite temperature, and of the form
\be\label{eq:Pi}
\Pi(\omega, q) \approx M_D^2 \frac{|\omega|}{v q}\,,
\ee
with $M_D^2 = \frac{g^2 k_F^2}{4\pi v N}$ the Debye scale. Combining (\ref{eq:Pi}) with (\ref{eq:L1}) leads, below $M_D$, to a boson with $z=3$ dispersion $q^3 \sim M_D^2 \omega$. The Debye scale is taken to be the UV cutoff of the low energy effective theory. A key simplification of working in the $z=3 $ scaling regime is that the fermion self-energy and gap depend predominantly on the frequency and not on the momentum.

The fermion self-energy is afflicted by thermal infrared singularities within perturbation theory. Ref. \cite{Wang:2017kab} showed how to resolve them by summing over rainbow diagrams. The result is the standard NFL term plus a new thermal term that is independent of frequency (other than a $\text{sgn}(\omega_n)$ required by causality),
\bea\label{eq:fullSigma}
\text{sgn}(\omega_n) \,\Sigma(\omega_n) &=& \Sigma_T + \Lambda_{\rm NFL}^{\epsilon/3} (2\pi T)^{1-\epsilon/3}\nonumber\\
&& \times\left[  \zeta(\epsilon/3)-\zeta(\epsilon/3,|n+1|)\right]\,.
\eea
where $\zeta$'s are Riemann zeta functions
\begin{align}
\zeta(x)=\sum_{n=1}^\infty \frac{1}{n^x},~~~\zeta(x,m)=\sum_{n=m}^\infty \frac{1}{n^x}.
\end{align}
We have a new scale
\be\label{eq:NFL}
{\Lambda_{\rm NFL}}\equiv {M_D} \left(\frac{1}{4\pi^2 \epsilon} \frac{g^2}{v }M_D^{-\epsilon}\right)^{3/\epsilon}\,,
\ee
below which NFL effects start to dominate.\footnote{Recall \cite{Raghu:2015sna} that in the $z=3$ regime, the engineering dimension $[g^2]=\epsilon$. So the right hand side in (\ref{eq:NFL}) is dimensionless.} These expressions hold only if $\Lambda_{\rm NFL}<M_D$. The thermal term is 
\be \label{eq:sigmaT}
\Sigma_T\approx\left(v^\epsilon\Lambda^{\epsilon/3}_{\rm NFL} M_D^{2\epsilon/3}\pi T\right)^{\frac{1}{1+\epsilon}}\,.
\ee

For $|n| \gg 1$ the difference of Riemann zeta functions asymptotes to 
\be
 (2\pi T)^{1-\epsilon/3}\left[  \zeta(\epsilon/3)-\zeta(\epsilon/3,|n+1|)\right] \approx |\omega_n|^{1-\epsilon/3}\,,
\ee
thus giving rise to the regular power-law behavior in the NFL state as is obtained at $T=0$. The fermion then acquires an anomalous dimension $\epsilon/3$. However, crucially, this difference $\zeta(\epsilon/3)-\zeta(\epsilon/3,|n+1|)$ vanishes at $n=-1,0$, i.e., $\omega_n=\pm \pi T$, and so the first Matsubara frequency is always dominated by the thermal term
\be
\Sigma(\pm \pi T) = \pm \Sigma_T\,.
\ee
We will analyze further the special role of the First Matsubara frequency in Sec.\ \ref{sec:mats}.

The emergence and resolution of the infrared singularity can be qualitatively understood by focusing on the $n=m$ term in the second line of \eqref{eq:SD},
\begin{align}
i\Sigma(\omega_n) =& -g^2 T \int \frac{d^d q}{(2\pi)^d}\frac{1}{q^2}\frac{1}{i\omega_n+i\Sigma(\omega_n)-v q_\perp}\nonumber\\
&+ \text{$m\neq n$ terms}\,.
\end{align}
Here we have linearized the dispersion relation around the Fermi surface, $\varepsilon(q)-\mu_F \approx v q_\perp$, and $q_\perp$ is the component of the momentum perpendicular to the Fermi surface. A widely used approach to solve this and other NFL integrals is to approximate the bosonic momentum by its component parallel to the Fermi surface, $q^2 \approx q_\parallel^2$. This is based on the Eliashberg approximation, which we review below in Sec.\ \ref{sec:beyondeliash}. However, this leads to a divergent integral over $q_\|$ for $d=3-\epsilon$. Instead, in this case we can perform the exact momentum integral to obtain
\bea
\sgn(\omega_n)\Sigma(\omega_n)&\approx &v^\epsilon\Lambda^{\epsilon/3}_{\rm NFL} M_D^{2\epsilon/3}\pi T \frac{1}{|{\omega_m+\Sigma(\omega_n)|^\epsilon}}\\
&+&  \Lambda_{\rm NFL}^{\frac{\epsilon}{3}} (2\pi T)^{1-\frac{\epsilon}{3}}\left[  \zeta(\frac{\epsilon}{3})-\zeta(\frac{\epsilon}{3},|n+1|)\right]\nonumber\,.
\label{eq:sethermal}
\eea
This gives a self-consistent equation for $\Sigma(\omega_n)$ that resums rainbow diagrams and resolves the IR singularity. To illustrate this, let us focus on the self-energy at the first Matsubara frequency: since the Riemann zeta functions cancel out, we arrive to an algebraic equation
\be
\Sigma(\pi T) =v^\epsilon\Lambda^{\epsilon/3}_{\rm NFL} M_D^{2\epsilon/3}\pi T \frac{1}{|\pi T+\Sigma(\pi T)|^\epsilon}\,,
\ee
whose solution reproduces (\ref{eq:sigmaT}) for temperatures below the NFL scale. The self-consistency of \eqref{eq:fullSigma} was checked in~\cite{Wang:2017kab}.

To summarize, the fermionic Green's function for low temperatures $\pi T \ll \Lambda_{\rm NFL}$ is characterized by
\be\label{eq:crossovers}
|\omega+\Sigma(\omega)| \approx \left \lbrace\begin{matrix}\omega\;, &  \omega \gg \Lambda_{\rm NFL}\\ \Lambda_{\rm NFL}^{ \epsilon/3} \omega^{1-\epsilon/3}\;,  & \Lambda_T \ll \omega \ll \Lambda_{\rm NFL}\\ \Sigma_T\;,& \omega \ll \Lambda_T \end{matrix}  \right.
\ee
with the crossover scale 
\be\label{eq:le}
\Lambda_T \sim \left( \Sigma_T \Lambda_{\rm NFL}^{-\epsilon/3}\right)^{\frac{1}{1-\epsilon/3}}.
\ee 
The finite $T$ quantum critical region then exhibits, in turn, Fermi liquid, quantum NFL, and ``thermal NFL" behaviors. The thermal term always dominates in the static limit, and violates the finite $T$ scaling of the QCP. We will see that it also plays an important role in the finite $T$ pairing problem.

%%%%%%%%%%%%%%%%%%%%%%%%%%%%%%%%%%%%%%
%%%%%%%%%%%%%%%%%%%%%%%%%%%%%%%%%%%%%%
\section{Non-Fermi liquid superconductivity}\label{sec:sc}

The interplay between SC and NFL behavior is encoded into the linearized gap equation
\bea\label{eq:gap1}
\t \Delta(\omega_n)&=& \frac{g^2}{N} \,\pi T_c \sum_m \int \frac{d^d q}{(2\pi)^d}\,D(\omega_n-\omega_m, q) \\
&\times& \frac{\t \Delta(\omega_m)}{(\omega_m+\Sigma(\omega_m))^2+ (\varepsilon(q)-\mu_F)^2}\nonumber\,,
\eea
where we have introduced the rescaled gap function 
\be
\t \Delta(\omega) \equiv Z(\omega) \Delta(\omega)\,.
\ee 
As we said, the factor 
$1/N$ shows the non-planar nature of SC discussed above, and $D(\omega,q)$ is the full boson propagator. In this section we will solve this equation, both numerically and analytically.

%%%%%%%%%%%%%%%%%
%%%%%%%%%%%%%%%%%
\subsection{Gap equation beyond the Eliashberg approximation}\label{sec:beyondeliash}

A standard approach to simplify (\ref{eq:gap1}) is to factorize the momentum integral between
parallel and perpendicular directions to the Fermi surface, assigning the perpendicular momentum dependence to the fermion Green's functions and the parallel one to the boson. This Eliashberg-type approximation is justified for nonzero bosonic frequency, as the boson has a larger dynamical exponent ($z_b=3$) than the fermion ($z_f=1-\epsilon/3$), and is much ``slower"---in analogy with Migdal's theorem for phonon mediated superconductivity. However, this procedure is problematic for zero bosonic frequency: it leads to a contribution
\begin{align}
\int q_\|^{1-\epsilon}dq_{\|} D(q_\|,|\omega_n-\omega_m|) \sim 1/|\omega_n-\omega_m|^{\epsilon/3}\,,
\end{align}
which diverges for the $m=n$ term in the sum, just like the case of the normal state.
 As in \cite{Wang:2017kab}, while factorization holds for $m \neq n$, it fails for the exchange of static bosons and has to be treated separately. 

As before, let us linearize the dispersion relation around the Fermi surface, $\varepsilon(q)-\mu_F \approx v q_\perp$, and parametrize $q_\perp= q \cos \theta$. We go beyond the factorization approximation by performing the full angular integration; for $\epsilon=3-d \ll 1$, this gives
\bea\label{eq:gap2}
\t \Delta(\omega_n) &=&\frac{g^2}{v N} \pi T_c \sum_m \int \frac{dq}{2\pi} \,q^{1-\epsilon} \,D(\omega_n-\omega_m,q) \nonumber \\
&&\times \tan^{-1} \left ( \frac{vq}{|Z(\omega_m) \omega_m|}\right)\,\frac{\t \Delta(\omega_m)}{| Z(\omega_m) \omega_m|}\,.
\eea
The Eliashberg approximation amounts to taking $\tan^{-1}(q/|Z\omega|) \sim \tan^{-1}(\omega^{1/3}/|Z\omega|) \to \pi/2$. But as we said, this leads to a singularity from exchange of zero frequency soft bosons.

The remaining calculation involves performing the $q$ integral for the $m \neq n$ and $m=n$ terms separately, and is presented in the Appendix \ref{app:1}. The final result is
\bea\label{eq:gap}
\t \Delta(\omega_n) &=& \Lambda_{\rm NFL}^{\epsilon/3} \frac{\pi T_c}{N} \Big \lbrace v^{\epsilon/3} M_D^{2\epsilon/3}\frac{\t \Delta(\omega_n)}{|\omega_n+\Sigma(\omega_n)|^{1+\epsilon}} \\
&+&  \sum_{m \neq n}\,\frac{1}{|\omega_m-\omega_n|^{\epsilon/3}}\,\frac{\t \Delta(\omega_m)}{|\omega_m+\Sigma(\omega_m)|}\Big\rbrace\nonumber\,.
\eea
The resolution of IR singularities then modifies the Eliashberg equation in two ways: there is a new diagonal term that resolves the $m=n$ exchange in the gap equation, and there is the contribution of the thermal NFL effects through $\Sigma(\omega)$. 

The diagonal term in the right hand side comes from the static part of the interaction, and is similar to disorder effects. Per Anderson's theorem, in a regular $s$-wave superconductor, disorder effects on the pairing gap and the self-energy cancel and do not affect $T_c$. This is not the case here -- however, this new diagonal term is suppressed by $1/N$ compared to the diagonal term in the self-energy. It is possible to redefine the gap function in order to remove this term, its effect being to reduce the thermal piece of the self-energy by $1-1/N\approx 1$  in the large $N$ limit. (For smaller $N$, the thermal term in the self-energy is effectively reduced and this could be numerically important.) See Appendix \ref{app:2} for more details. Thus the diagonal term in the gap equation can be safely discarded. On the other hand, we will see that the thermal term in the self-energy plays a crucial role in the dynamics of SC.

Let us first solve (\ref{eq:gap}) numerically and then develop analytic methods. Since $T_c$ is approached from above, the problem amounts to finding the largest eigenvalue and eigenvector of the kernel that appears in the right hand side of (\ref{eq:gap}). We do this by fixing a maximum Matsubara frequency $\omega_{max}=(2 M_{max}+1) \pi T_c$, and then increase $M_{max}$ to check for convergence. Furthermore, for $\omega_n \gg \pi T_c$ it is sufficient to sample the Matsubara frequencies on exponential intervals, which allows to access parametrically small $T_c$. 

\begin{figure}[h]
\begin{center}
\includegraphics[width=\columnwidth]{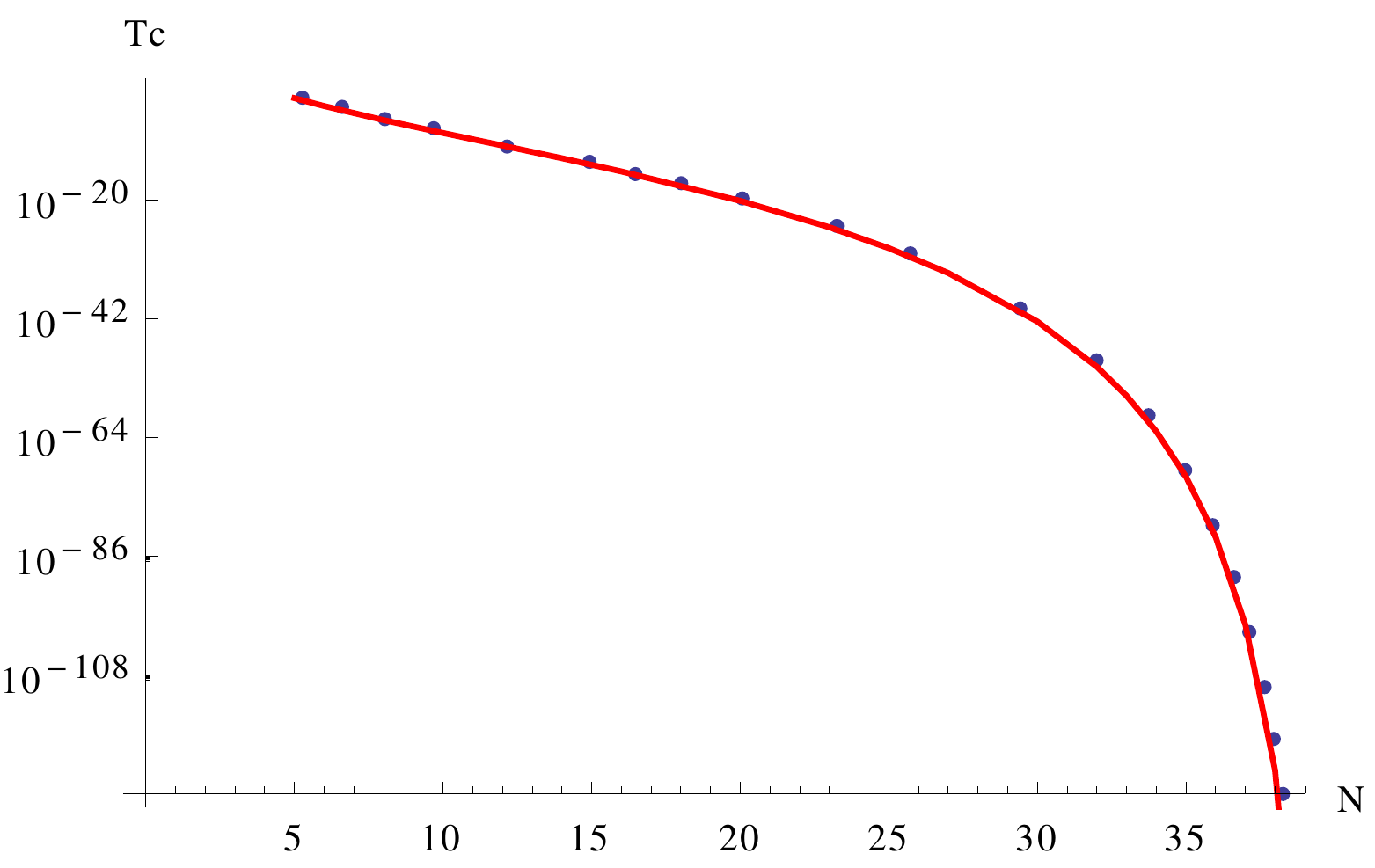} 
\caption{Blue dots: critical temperature as a function of $N$, for $\epsilon=0.3, \Lambda_{\rm NFL}/M_D=1$. Red curve: prediction from the differential equation (\ref{eq:Ddiff}). A BKT-like scaling for $T_c=0$ is observed at $N=40$. }
\label{fig:Tc}
\end{center}
\end{figure}

Fig. \ref{fig:Tc} shows $T_c$ as a function of $N$. It decreases for larger $N$, and goes to zero smoothly for $N> N_{c}\approx 40$ for the present choice $\epsilon=0.3$. Further numerical analysis reveals that higher order derivatives also vanish smoothly, thus suggesting a Berezinskii-Kosterlitz-Thouless (BKT) scaling for $T_c(N)$. In fact, a similar behavior for the SC gap at zero temperature was found in \cite{Raghu:2015sna, Wang:2016hir}. To test this further, we need to access exponentially small temperatures, which requires increasing $N_{max}$ and is thus numerically costly. For this reason we turn now to approximate analytical approaches. 

%%%%%%%%%%%%%%%%%
%%%%%%%%%%%%%%%%%
\subsection{Local frequency approximation}

Since we have taken care of the infrared divergence at zero bosonic frequency ($m=n$), it is consistent to replace the Matsubara sum by an integral,
\be\label{eq:Deltaint}
\t \Delta(\omega')=\frac{1}{2N} \int_{|\omega|>\pi T_c} d\omega\, u(\omega-\omega')\,\frac{\t \Delta(\omega)}{|\omega +\Sigma(\omega)|}\,,
\ee
where the lower boundary $|\omega|>\pi T_c$ encodes the discreteness of Matsubara sum, and we have introduced the kernel 
\be
u(\omega)\equiv\left(\frac{\Lambda_{\rm NFL}}{|\omega|}\right)^{\epsilon/3}\,.
\ee 
As in \cite{Wang:2016hir}, (\ref{eq:Deltaint}) can be transformed into a differential equation by a \textit{local} approximation\footnote{A similar approximation was first used in \cite{Son:1998uk}.}
\be\label{eq:ulocal}
u(\omega-\omega') \approx \left\lbrace \begin{matrix} u(\omega') &,&|\omega|< |\omega'| \\ u(\omega) &,&|\omega|> |\omega'|\end{matrix} \right. \,.
\ee
The intuition behind this approximation comes from the Wilsonian RG, where quantum corrections at a given scale $\omega$ are dominated by virtual loops with characteristic energy of order $\omega$. Now, here we are evaluating a 1PI quantity, the gap function, so we need to check that a Wilsonian approximation holds. We have checked this by comparing with numerical results, finding excellent agreement; we give an example below. Ultimately, the validity of this approach rests on the fact that $\epsilon$ is small.\footnote{For $\epsilon \sim 1$, the regime of $u(\omega-\omega')$ with $\omega \sim \omega'$ becomes more important. This range is not taken into account by (\ref{eq:ulocal}).}

The change of variables
\be\label{eq:change}
\omega = \Lambda_{\rm NFL} e^{-3 x/\epsilon}
\ee
then yields the differential system
\bea\label{eq:Ddiff}
&&\t  \Delta''(x) -\t  \Delta'(x) + \frac{g_1 e^x}{ Z(x)} \,\t \Delta(x) =0 \nonumber\\
&&\t  \Delta'(x_c) = 0\;\;,\;\;\t \Delta(x \to -\infty) =0\,.
\eea
Here $\pi T_c = \Lambda_{\rm NFL} e^{-3 x_c/\epsilon}$, and
\be
Z(x)=1+\frac{\Sigma(\omega)}{\omega}= 1+e^x+e^{x_T} e^{3(x-x_T)/\epsilon}\, ,\;g_1 \equiv \frac{3}{\epsilon N}\,,
\ee
and $x_T$ is defined via $\Lambda_T = \Lambda_{\rm NFL} e^{-3 x_T/\epsilon}$, namely the thermal NFL scale $\Lambda_T$ of (\ref{eq:crossovers}) in the $x$-variables.
The term $e^x$ in $Z(x)$ encodes the zero temperature NFL self-energy, while the last term comes from the thermal term $\Sigma_T$. This equation should be solved for different values of $x_T$ (itself a function of $x_c$ or $T_c$) until the IR boundary condition $\t \Delta'(x_c)=0$ is satisfied. This determines the critical temperature $T_c$.

Eq.~(\ref{eq:Ddiff}) describes all possible behaviors of the SC gap at low temperatures. For $x<0$ (i.e. $\omega > \Lambda_{\rm NFL}$), $Z(x) \approx 1$ and NFL effects from the self-energy are negligible. The solution is a Bessel function, a result which reproduces the gap behavior of color superconductivity found in \cite{Son:1998uk}. The quantum NFL term starts to become important around $x=0$, and $Z(x) \approx 1+e^x$.
At $T \neq 0$, $x_T$ is finite, and the last term in $Z(x)$ quickly dominates for $x>x_T$ (i.e. $\omega<\Lambda_T$). When this occurs, $\t \Delta$ approaches a constant -- the energy scale below which the thermal term dominates acts like a gap that stops the growth of the SC instability.  Summarizing, we obtain 
\be\label{eq:Deltaregimes}
\t\Delta(x) \approx \left \lbrace\begin{matrix}e^{x/2} J_1(2 \sqrt{g_1} e^{x/2})\;, &  x<0\\ c_1 e^{x/2} \exp{\(ix \sqrt{g_1-\frac{1}{4}}\)}+c.c.\;,  & 0 < x < x_T \\ c_2\;,& x> x_T \end{matrix}  \right. \,.
\ee
This is the solution from keeping only a single term of $Z(x)$ that dominates in each range, and $c_1, c_2$ are determined by smoothly connecting across the intervals. In terms of the original $\omega$ variable, the gap function behaves as
\be\label{eq:Deltaregimes2}
\t\Delta(\omega) \sim \left \lbrace\begin{matrix}\omega^{-\epsilon/3}\;, &  \omega>\Lambda_{\rm NFL}\\  \omega^{-\epsilon/6} \cos(\frac\epsilon 3\sqrt{g_1-\frac{1}{4}}\ln \omega+\phi)\;,  & \Lambda_T < \omega < \Lambda_{\rm NFL} \\ \rm{const}\;,& \omega> \Lambda_T \end{matrix}  \right. \,,
\ee
where $\phi$ is a phase.

\begin{figure}[h]
\begin{center}
\includegraphics[width=\columnwidth]{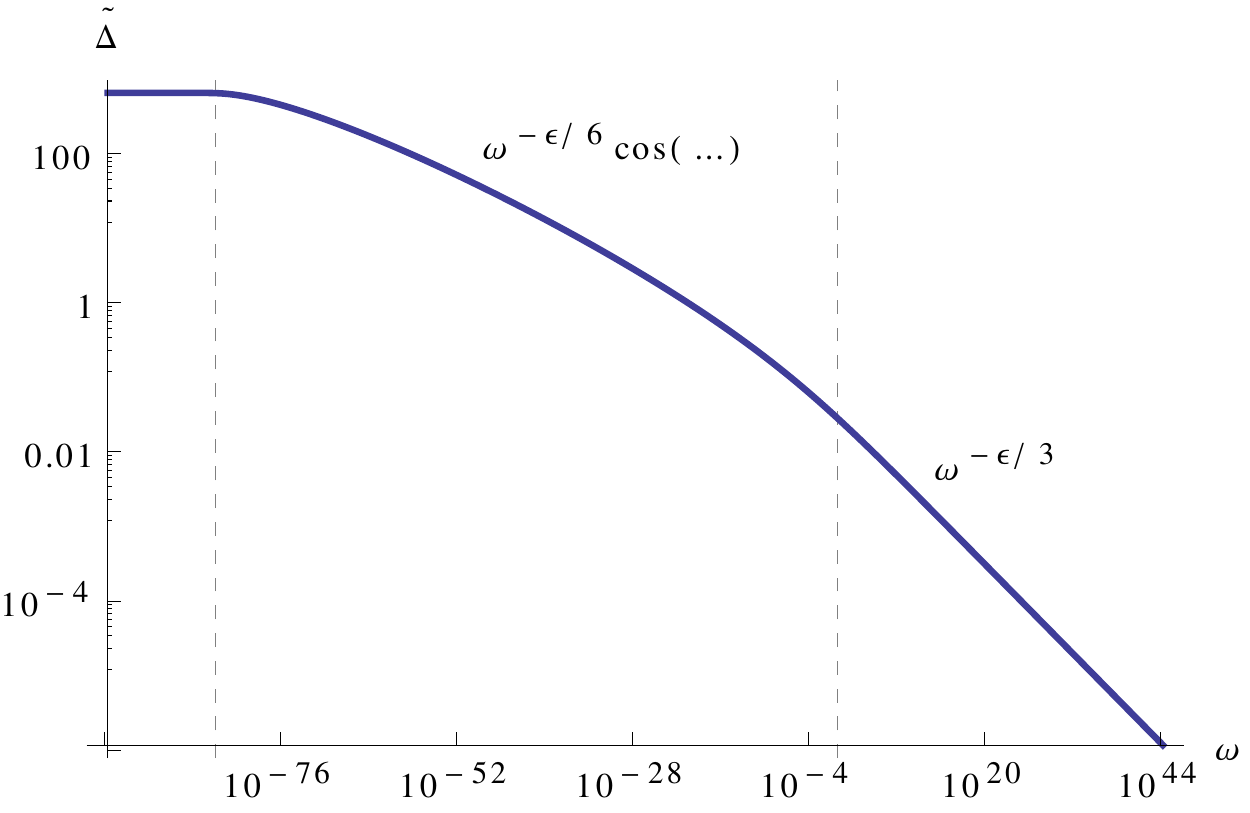} 
\caption{Gap eigenvector from the full numerical solution. The result from the differential equation approximation completely overlaps within the plotting accuracy. Parameters: $T_c=10^{-100}M_D, \Lambda_{\rm NFL}=M_D, N=37.1, \epsilon=0.3$. }
\label{fig:gap}
\end{center}
\end{figure}

By matching the last two regimes in \eqref{eq:Deltaregimes}, we find that the dominant effect of finite temperature is simply to modify the IR boundary condition to 
\be\label{eq:newIR}
\t \Delta'(x_T)=0\,,
\ee
because $\t \Delta$ approaches a constant for $x> x_T$, i.e., $\omega_m<\Lambda_T$. This is a direct consequence of the thermal NFL regime in the self-energy.  The behavior of $\t \Delta(\omega)$ is illustrated in Fig. \ref{fig:gap}, showing an excellent agreement between the solutions to (\ref{eq:gap}) and (\ref{eq:Ddiff}). 

At this point it is instructive to compare the finite-$T$ and zero-$T$ pairing problems. Under the same local approximation, the pairing problem at $T=0$ can similarly be transformed into an integral equation in the same form as \eqref{eq:Ddiff} for $\omega >\Delta_0\equiv\Delta(T=0)$~\cite{Wang:2016hir}. The main difference is that there are no thermal effects, and hence $Z(x)=1+e^x$. The solution for $\tilde\Delta$ only has two different behaviors---the first two ranges in \eqref{eq:Deltaregimes}---and the boundary condition \eqref{eq:newIR} is replaced by $\t \Delta'(\omega=\Delta_0)=0$. We note that for $T=0$ the differential equation and its IR boundary condition can also be derived using RG beta function of the BCS 4-Fermi coupling~\cite{Wang:2016hir}. Hence by comparing with (\ref{eq:newIR}) we deduce that 
\be\label{eq:corresp}
\Lambda_{T_c} \approx \Delta_0\,.
\ee

Let us now compare the SC gap at zero temperature with $T_c$ obtained here. While one would expect $\Delta_0 /T_c\approx 1.76 = O(1)$, e.g. as in BCS theory, this is not what happens for NFLs. Instead, from $\Delta_0 \sim \Lambda_{T_c}$ and Eq.\ \eqref{eq:le}, we have
\bea\label{eq:Delta0Tc}
\frac{\Delta_0}{T_c}&\sim& \left[v^3\left(\frac{M_D}{\Lambda_{\rm NFL}} \right)^\epsilon 
 \left(\frac{M_D}{T_c} \right)^{2-\epsilon} \right]^{\frac{\epsilon}{(3-\epsilon)(1+\epsilon)}}\nonumber\\
 & \approx &  \left(\frac{M_D}{T_c} \right)^{2\epsilon/3}\gg 1\,.
\eea
The thermal fluctuation then parametically supresses $T_c$ and generates two different scales at zero and finite temperature. (A scenario in which thermal fluctuations modestly suppress $T_c$ was discussed in Ref.\ \cite{acn}.) The relation \eqref{eq:Delta0Tc} is one of the central results of this work. We comment that indeed in many unconventional superconductors the ratio $\Delta_0/T_c$ is large~\cite{dtc-matsuda,dtc-cuprates,dtc-hhwen}. This is typically attributed to fluctuations of the SC order at finite temperature~\cite{emery-kivelson}.
Our work points to the role of thermal fluctuations as an alternative possible scenario for a large $\Delta_0/T_c$.

As we discussed, the critical temperature $T_c(N)$ is determined by matching the boundary conditions. A careful examination  of \eqref{eq:Deltaregimes} shows that this is only possible for $g_1\equiv 3/{\epsilon N}>1/4$, or \be
N>N_c=12/\epsilon,
\ee
to ensure non-monotonic (oscillatory) behavior~\cite{acf,Wang:2016hir} in $0<x<x_T$ that can be used to match the functional form in other regimes at both ends.
 In physical terms, NFL effects are subdominant for $N \epsilon \ll 12$, but start to strongly decrease $T_c$ as $N \epsilon$ tends to $12$. Matching the values and slopes of the gap function in different regimes (\ref{eq:Deltaregimes}) near $N_c$ obtains
\be\label{eq:T_c}
T_c \sim \exp \left[-\left(\frac{3}{\epsilon}-1\right)(1+\epsilon) \frac{2\pi}{\sqrt{\frac{12}{\epsilon N}-1}} \right]\,\Lambda_{\rm NFL}\,,
\ee
which exhibits BKT scaling as $N \to N_c$.
The relation between $T_c$ and $N$ obtained with this method is shown in the red curve of Fig.~\ref{fig:Tc}, and the agreement with (\ref{eq:gap}) is excellent. The same expression follows from the correspondence (\ref{eq:corresp}) with the $T=0$ problem.

The BKT scaling for the superconducting temperature in the NFL regime appears surprising at first. Indeed, the physics here is quite different from that of a classical BKT transition driven by the unbinding of vortices. However, as shown in~\cite{Kaplan:2009kr}, BKT scaling arises quite generically from the collision of fixed points in the RG, and this is also the underlying mechanism operating in our setup. To see how this comes about, consider $N> N_c$, so that $T_c=0$. In this case, SC is completely extinguished down to the lowest temperatures, and we can use the $T=0$ analysis of~\cite{Raghu:2015sna}. This shows that the RG beta function for the BCS 4-Fermi coupling $\lambda_{BCS}$ admits two fixed points. One fixed point is stable and leads to a QCP with critical pairing fluctuations; the other fixed point is unstable, namely it arises as a high energy critical point. As $N \to N_c$ from above, these fixed points annihilate each other. This is the physics responsible for the BKT scaling we have observed, however, as stressed before, the anomalous scaling between $T_c$ and the gap (which is also the dynamical scale generated by the RG) is a purely thermal effect.

To summarize, for $N>N_c$ we enter a quantum critical state that corresponds to the finite temperature regime of a QCP with critical pairing fluctuations and where the superconducting instability is irrelevant. The properties of this normal state was recently discussed in~\cite{Wang:2017kab}. Our analysis thus establishes a $(T,N)$ phase diagram (Fig.\ \ref{fig:Tc}) where the SC phase lies below the red curve.

%%%%%%%%%%%%%%%%%
%%%%%%%%%%%%%%%%%
\subsection{Approximate solution to the integral equation}

It turns out that the above approximate result can also be obtained by directly solving the integral equation. The method is similar to that used in ~\cite{acf,ywfuture}. To proceed, we focus on external frequencies $\Lambda_{T_c}\ll \omega'\ll \Lambda_{\rm NFL}$, and consider the SC gap that is even in frequency $\t \Delta(\omega)=\t \Delta(-\omega)$. We can then break the integral equation (\ref{eq:Deltaint}) into the following pieces:
\begin{widetext}
\be\label{eq:integral_break}
\Lambda_{\rm NFL}^{-\epsilon/3}\t\Delta(\omega')=\frac{1}{N}\[\int^{\Lambda_{T_c}}_{\pi T_c}\frac{d\omega}{\left(\omega'\right)^{\epsilon/3}}\frac{\t\Delta(\omega)}{\Sigma_{T_c}}
+\int^{\omega'}_{\Lambda_{T_c}}\frac{d\omega}{\left(\omega'\right)^{\epsilon/3}}\frac{\t\Delta(\omega)}{\Lambda_{\rm NFL}^{\epsilon/3}\omega^{1-\epsilon/3}}
+\int^{\Lambda_{\rm NFL}}_{\omega'}d\omega \frac{\t\Delta(\omega)}{\Lambda_{\rm NFL}^{\epsilon/3}\omega}+\frac{1}{N}\int^{\Lambda_{UV}}_{\Lambda_{\rm NFL}}d\omega \frac{\t\Delta(\omega)}{\omega^{1+\epsilon/3}}\]
\ee
\end{widetext}
\begin{comment}
\bea\label{eq:integral_break}
&&\Lambda_{\rm NFL}^{-\epsilon/3}\Delta(\omega')=\frac{1}{N}\int^{\Lambda_{T_c}}_{\pi T_c}d\omega\left(\omega'\right)^{-\epsilon/3}\frac{\Delta(\omega)}{\Sigma_{T_c}}\nonumber\\
&+&\frac{1}{N}\int^{\omega'}_{\Lambda_{T_c}}d\omega\left(\omega'\right)^{-\epsilon/3}\frac{\Delta(\omega)}{\Lambda_{\rm NFL}^{\epsilon/3}\omega^{1-\epsilon/3}}\nonumber\\
&+&\frac{1}{N}\int^{\Lambda_{\rm NFL}}_{\omega'}d\omega \frac{\Delta(\omega)}{\Lambda_{\rm NFL}^{\epsilon/3}\omega}+\frac{1}{N}\int^{\Lambda_{UV}}_{\Lambda_{\rm NFL}}d\omega \frac{\Delta(\omega)}{\omega^{1+\epsilon/3}}\nonumber
\eea
\end{comment}
where we have approximated  $\omega+\Sigma(\omega)$ by its various contributions, and $u(\omega-\omega')$ using the local approximation. 

The strategy is to solve self-consistently for a power-law solution $\t\Delta(\omega)\sim A\omega^\gamma$ in this regime, where the paring is controlled by the QCP. We assume and verify later that the SC instability is dominated by the frequency range $\Lambda_{T_c}\ll \omega'\ll \Lambda_{\rm NFL}$ (as we have seen in the last subsection); the feedback effects from other frequency ranges are parametrically small. This leads to
\bea\label{eq:integral_Eq}
&& N A\left(\omega'\right)^\gamma=A\left(\frac{1}{\gamma+\epsilon/3}-\frac{1}{\gamma}\right)\left(\omega'\right)^\gamma + K\left(\omega',T_c\right)\nonumber\\
&& K\left(\omega',T_c\right)=\frac{A}{\gamma}\Lambda_{\rm NFL}^\gamma-\frac{A \Lambda_{T_c}^{\gamma+\epsilon/3}}{\gamma+\epsilon/3}\left(\omega'\right)^{-\epsilon/3}\,.
\eea
Observe that without the second term $K(\omega',T_c)$, power-law solutions exist for any $\epsilon$ and $N$: 
\be
\gamma = -\frac{\epsilon}{6}\left(1\pm 2i\beta\right),\;\beta = \sqrt{\frac{3}{\epsilon N}-\frac{1}{4}},
\ee
so finding $T_c$ is equivalent to the condition 
\be\label{eq:bdrycond}
\Re \, {K}(\omega', T_c)=0,~\text{for any } \omega'.
\ee
Here it suffices for the real part of $K$ to vanish because the equation is linear and the integral kernel is real, and real and imaginary parts of $\tilde\Delta\sim \omega^\gamma$ decouple as two independent candidate solutions. 
The condition \eqref{eq:bdrycond} is only possible for a complex exponent $\gamma$, in which case $\Re \,K$ as a function of $\Lambda_{\rm NFL}$ and $\Lambda_T$ is oscillatory and has zeros. Then  \eqref{eq:bdrycond} can be satisfied by tuning the phase of the complex amplitude $A\sim e^{i\phi}$ and $T_c$ (or $\Lambda_{T_c}$). Requiring complex $\gamma$ reproduces the SC condition $N<N_c\equiv 12/\epsilon$; near the phase transition $\beta \to 0$, and solving $\text{Re}\, K(\omega',T_c)=0$ requires that
\bea\label{eq: self-consistency}
\phi +\frac{\epsilon}{3}\beta \ln{\Lambda_{\rm NFL}}&=&\left(\frac{1}{2}+m\right)\pi+\mathcal{O}(\beta),\;m\in\mathbb{Z}\nonumber\\
\phi +\frac{\epsilon}{3}\beta \ln{\Lambda_{T_c}}&=&\left(\frac{1}{2}+n\right)\pi+\mathcal{O}(\beta),\;n\in\mathbb{Z}\nonumber\,.
\eea 
Eliminating $\phi$ and picking the maximum $\Lambda_{T_c}\leq \Lambda_{\rm NFL}$ reproduces (\ref{eq:T_c}). 

The functional behavior of  $\t\Delta(\omega')$ in the other regimes can be easily determined by performing an integral over $\Lambda_{T_c}\ll \omega'\ll \Lambda_{\rm NFL}$, which only involves the known $\t\Delta(\omega')$, and does not require solving integral equation.  This gives
\be
\t \Delta(\omega')\sim \begin{cases} 
\left(\Lambda_{\rm NFL}\right)^{\gamma+\epsilon/3}\left(\omega'\right)^{-\epsilon/3},\; \;\;\;\;\;\;\omega'\gg \Lambda_{\rm NFL}\\
(\omega')^{\gamma},\;\;\;\;\;\;\; \;\;\;\;\;\;\;\;\;\;\;\;\;\;\;\Lambda_{T}\ll\omega'\ll\Lambda_{\rm NFL}\\
\left(\Lambda_{T_c}\right)^{\gamma},\;\;\;\;\;\;\;\;\;\;\;\;\;\;\;\;\;\;\;\pi T_c \ll \omega' \ll \Lambda_{T_c}\\
\end{cases} 
\ee
in agreement with (\ref{eq:Deltaregimes2}).
The resulting piece-wise solution $\t \Delta(\omega')$ is smooth. Furthermore, it can be verified that feedback effects in \eqref{eq:integral_break} from regimes $\omega\ll \Lambda_{T_c}$ and $\omega\gg \Lambda_{\rm NFL}$ are negligible. This solution is therefore self-consistent.

%%%%%%%%%%%%%%%%%%%%%%%%%%%%%%%%%%%%%%
%%%%%%%%%%%%%%%%%%%%%%%%%%%%%%%%%%%%%%
\section{Revisiting the role of the first Matsubara frequency}\label{sec:mats}

The result we obtained has found that near the quantum critical region and for $N>12/\epsilon$, SC is avoided due to (thermal and quantum) NFL dynamics. On the other hand, the work~\cite{2016PhRvL.117o7001W} argued that SC always persists at finite $T$ due to the special role of the first Matsubara frequency, at which the NFL behavior is absent and hence SC is strongly enhanced. We emphasize here that this ``Fermi-liquid first Matsubara physics'' is based on the cancellation of thermal contributions to the gap equation and the self-energy, which in our model is only realized at $N=1$ (see Appendix \ref{app:2}). For $N>1$, the thermal piece in the NFL self-energy dominates at lowest frequencies, and the FL first Matsubara physics is spoiled. 

To see how exactly the first Matsubara physics interplays with thermal fluctuations, 
it is interesting to modify the previous theory so that thermal fluctuations can be made parametrically small. In this case we show that the first Matsubara physics can partially re-emerge.

Let's rewrite the frequency dependent part in the fermionic kernel for $\omega_m>0$ as [see (\ref{eq:fullSigma})]
\bea
\omega_m+\Sigma(\omega_m)&=& |\omega_m|+\(\lambda\pi T \)^{\frac{1}{1+\epsilon}}\nonumber\\
&+&\Lambda_{\rm NFL}^{\epsilon/3}(2\pi T)^{1-\epsilon/3}\left(\zeta(\epsilon/3)-\zeta(\epsilon/3, m+1)\right)\nonumber\\
&\approx& |\omega_m|+(\lambda\pi T)^{\frac{1}{1+\epsilon}}
+\Lambda_{\rm NFL}^{\epsilon/3}|\omega_n|^{1-\epsilon/3}
\label{crossover2}
\eea
where we introduced a new parameter $\lambda\equiv v^{2\epsilon/3}\Lambda_{\text{NFL}}^{\epsilon/3}M_D^{2\epsilon/3}$ and for convenience we have set $M_D=1$. The approximate result in the last line is for $m\gg 1$. However,  as we mentioned, the first Matsubara frequency is special: at $m=0$, the last term vanishes, and only the thermal piece contributes to the self-energy, $\Sigma(\pi T)=\(\lambda\pi T \)^{\frac{1}{1+\epsilon}}$.

We now take the ``artificial" limit of small $\lambda$ and analyze the special role of the first Matsubara frequency. Consider the possibility that the gap equation (\ref{eq:gap}) can be truncated at $\omega_n = \pm \pi T$. This truncation at the first level is consistent if the contributions $\Delta(\omega_n)$ from higher frequencies are small. This requires
\be\label{eq:highneglect}
\Lambda_{\rm NFL}^{\epsilon/3}|\omega_n|^{1-\epsilon/3} \gg \pi T_c + \Sigma(\pi T_c)\,. 
\ee This gives an algebraic equation for the SC gap
\be
\Delta(\pi T_c) \approx \Lambda_{\rm NFL}^{\epsilon/3} \frac{\pi T_c}{N} \, \frac{1}{(2\pi T_c)^{\epsilon/3}}\,\frac{\Delta(-\pi T_c)}{\pi T_c + \Sigma(\pi T_c)}\,.
\ee
The values of the gap at higher frequencies are determined iteratively in terms of $\Delta(\pi T_c)$.
Since the gap function is symmetric, we arrive at
\be\label{eq:NfirstTc}
N = \left( \frac{\Lambda_{\rm NFL}}{2\pi T_c}\right)^{\epsilon/3}\left(1+\frac{\Sigma(\pi T_c)}{\pi T_c}\right)^{-1}\,.
\ee
Remember that (\ref{eq:NfirstTc}) is valid only if (\ref{eq:highneglect}) is satisfied, which will eventually be violated for low enough $T_c$. This implies that there exist different regimes. Let us discuss them in details:
\vskip 2mm
\noindent $\bullet$ $\lambda^{1/\epsilon}\ll T_c\ll \Lambda_{\rm NFL}$. This range for $T_c$ corresponds to requiring (\ref{eq:highneglect}) and $\pi T_c \gg \Sigma(\pi T_c)$. From (\ref{eq:NfirstTc}), we have
\be
N\approx \left(\frac{\Lambda_{\rm NFL}}{2\pi T_c} \right)^{\epsilon/3}\,.
\ee
This automatically gives $T_c \ll \Lambda_{\rm NFL}$ for any $N>1$, which then ensures (\ref{eq:highneglect}). Pairing is driven by the first Matsubara effect and is Fermi-liquid like; the NFL behavior dominates for all other frequencies, and heir contribution in the pairing equation is parametrically smaller. This reproduces in our model the results found in~Ref.\ \onlinecite{2016PhRvL.117o7001W}.
For this result to be compatible with the temperature range we are in, we need 
\be\label{eq:Ncond}
N \ll N_c'\equiv (\Lambda_{\rm NFL}^\epsilon \lambda^{-1})^{1/3}\,.
\ee

%\vskip 1mm

\noindent $\bullet$ $(\lambda^{1/\epsilon}\Lambda_{\rm NFL}^{-1})^{\frac{1+\epsilon}{2-\epsilon}}\lambda^{1/\epsilon}\ll T_c\ll \lambda^{1/\epsilon}$. This range for $T_c$ corresponds to requiring (\ref{eq:highneglect}), but $\Sigma(\pi T_c) \gg \pi T_c$. This is then a new NFL regime, where pairing is still driven by the first Matsubara frequency, but the solutions are now driven by the thermal term instead of the bare frequency term. From (\ref{eq:NfirstTc}), one obtains a different scaling for $N(T_c)$:
\be
N \approx (\Lambda_{\rm NFL}^\epsilon \lambda^{-1})^{\frac{1}{1+\epsilon}} \left(\frac{T_c}{\Lambda_{\rm NFL}} \right)^{\frac{2\epsilon-\epsilon^2}{3+3\epsilon}}
\ee
In this range $T_c$ increases upon increasing $N$. Compatibility with the temperature range requires the same relation (\ref{eq:Ncond}) to hold.
 
 %Interestingly, for a fixed $N$, when T is {\it increased} from $T_c$, the SC gap equation starts to develop nonlinear solutions, in contrast to usual superconducting transitions.}
\vskip 1mm

\noindent$\bullet$ $T_c\ll (\lambda^{1/\epsilon}\Lambda_{\rm NFL}^{-1})^{\frac{1+\epsilon}{2-\epsilon}}\lambda^{1/\epsilon}$. In this regime, the thermal term dominates all the lowest Matsubara frequencies (not just the first one), in other words, $\Sigma(\pi T_c)\gg \Lambda_{\rm NFL}^{\epsilon/3}|\omega_n|^{1-\epsilon/3}$. The special role of the first Matsubara frequency is thus lost. The pairing is driven by the NFL dynamics, i.e., the interplay between  the thermal and $z=3$ contributions of the boson to the fermionic self-energy that we analyzed in the previous section: 
\be
N(T_c\to 0) \to N_c=12/\epsilon,
\ee
\emph{independent} of $\lambda$. The independence of $N(T_c \to 0)$ on $\lambda$ is because $\lambda$ only affects the scale $\Lambda_T$, which tends to zero anyways at $T\to 0$.

\begin{figure}[h]
 \includegraphics[width=\columnwidth]{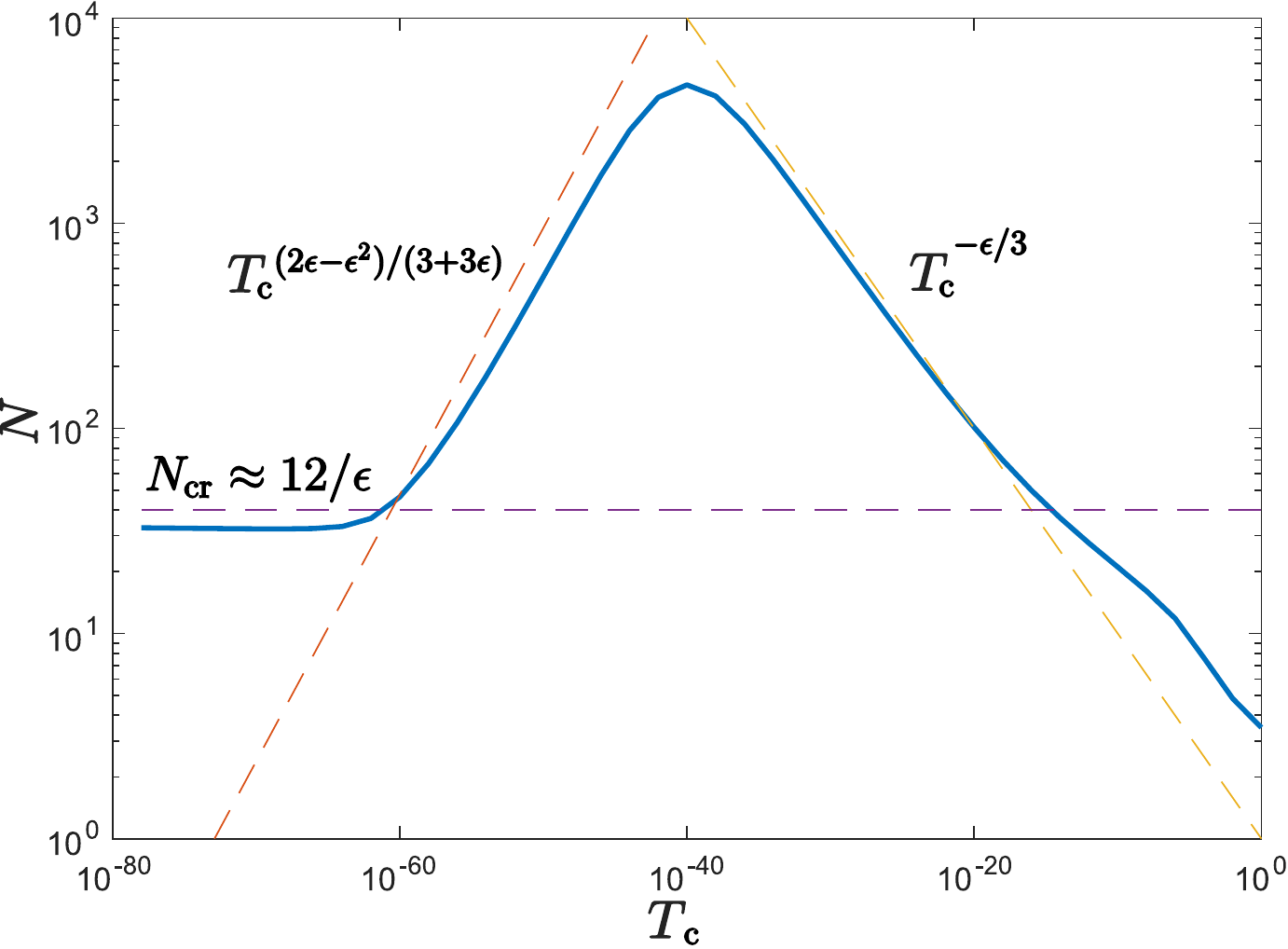}
 \caption{The critical $N$ as a function of temperature $T$ for $\epsilon=0.3$, and $\lambda=10^{-12}$.}
 \label{fig:fig2}
 \end{figure}

We conclude that the FL first Matsubara physics of~\cite{2016PhRvL.117o7001W} dominates at higher $T$, while the NFL-driven first Matsubara behavior emerges at lower $T$. SC is enhanced in the FL regime, while thermal and quantum NFL fluctuations suppress it. This suggests the intriguing possibility of a ``re-entrance behavior'' -- an SC phase at high $T$ and a return to the NFL normal state at lower temperatures.

To explore this further, let us connect the three ranges we just obtained.
For $N\gg N_c'\equiv (\Lambda_{\rm NFL}^\epsilon \lambda^{-1})^{1/3}$, there is no solution for $T_c$ and this corresponds to the NFL normal state. For $N_c\equiv 12/\epsilon\ll N\ll N_c'$, there are two solutions $T_{c1,c2}$, between which one can show that a nonzero SC gap develops. These temperatures then correspond to entry/exit critical temperatures for the re-entrant SC phase.
Finally, for $N<N_c$, SC develops at a finite temperature and persists down to zero temperature. We have verified this numerically, solving the gap equation for  a small parameter $\lambda$ in the thermal piece of the self-energy. We show the $N=N(T_c)$ curve in Fig.~\ref{fig:fig2}, for $\epsilon=0.3$ and $\lambda=10^{-12}$. 
 
 \begin{figure*}
 \includegraphics[width=1.6\columnwidth]{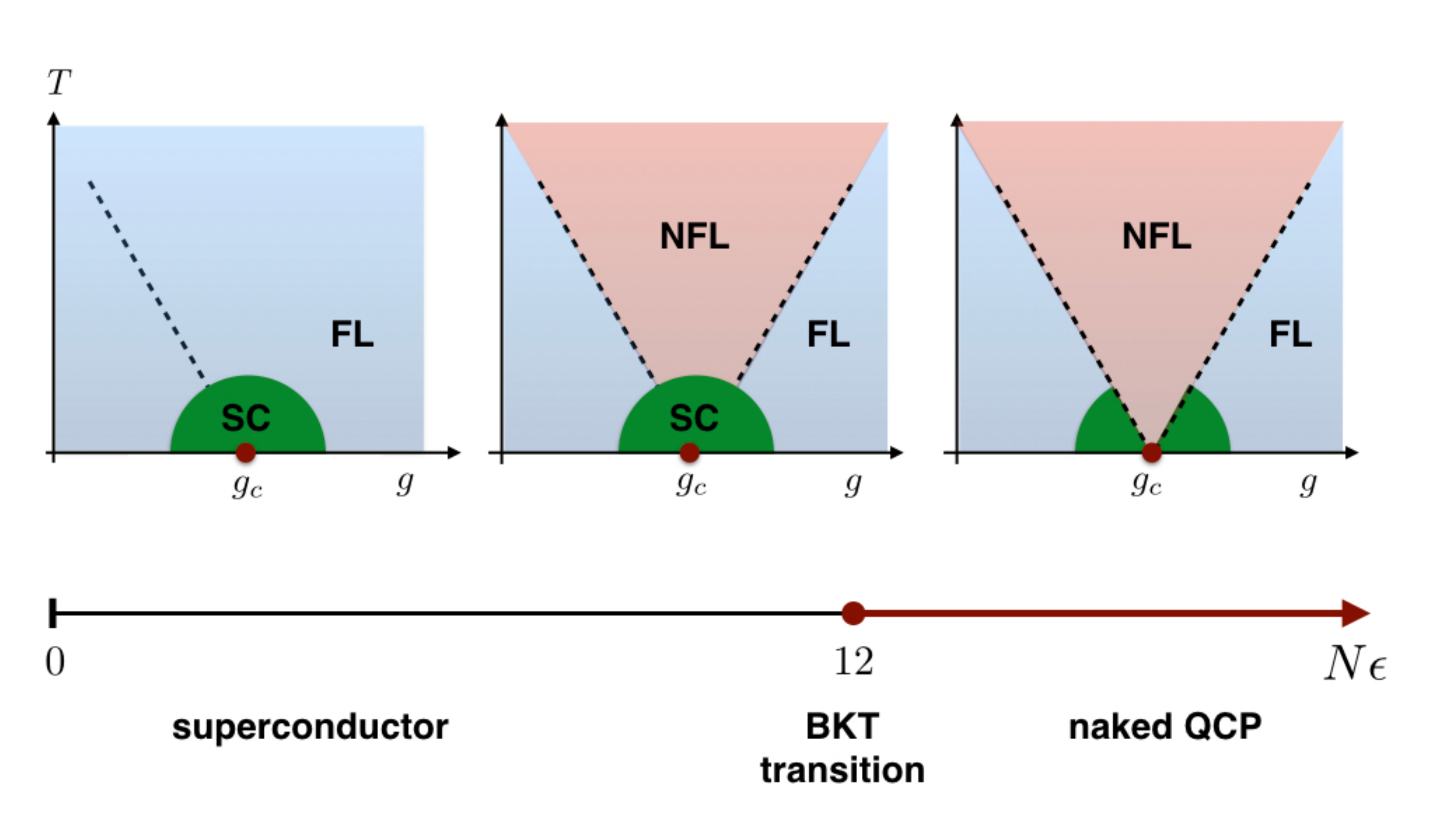}
 \caption{Phase structure for different values of $N \epsilon$ near the QCP at $g=g_c$ (a parameter that tunes the boson to criticality).}
 \label{fig:phases}
 \end{figure*}

So far we have used the external parameter $\lambda$ to access the regimes where the first Matsubara frequency plays a leading role. However, it is important to determine whether $\lambda \Lambda_{\rm NFL}^{-\epsilon} \ll 1$ can be physically realized in the class of models studied here. Comparing \eqref{crossover2} and (\ref{eq:sigmaT}) gives
\be\label{eq:lambdasmall}
\lambda = \left(v \frac{M_D}{\Lambda_{\rm NFL}} \right)^{2 \epsilon/3} \Lambda_{\rm NFL}^{\epsilon}\,,
\ee
so it is tempting to identify the ``strong-coupling" limit $\Lambda_{\rm NFL}\gg M_D$ as a realization of the small parameter $\lambda \Lambda_{\rm NFL}^{-\epsilon} \ll 1$. However, we caution that in this limit the formulas for the self-energy and the gap are no longer valid, since the $z=3$ scaling of the boson breaks down. Another possibility would be to take $v \ll 1$, so that the fermion is much slower than the boson. But in this case the $z=3$ Landau-damped boson emerges only for $\omega \ll v M_D$~\cite{Torroba:2014gqa}. Then $v M_D$ replaces $\Lambda_{\rm NFL}$ as the new UV cutoff, and (\ref{eq:lambdasmall}) cannot be made small. We conclude that in the regime of parametric control that has been our focus, the special role of the first Matsubara frequency (both in its FL and NFL versions) is lost, and SC is driven by the NFL dynamics of Sec.~\ref{sec:sc}.

Nonetheless, the re-entrant behavior cannot be completely excluded --- it could arise for instance in the intermediate undamped regime where the self-energies depends both on frequency and momenta. Further numerical studies are desired to shed more light on this issue. On the other hand, it would be extremely interesting to find a modification of our model that realizes the small $\lambda$ parameter in a controlled way, for example via disorder, or changing the dimension of the problem.

%%%%%%%%%%%%%%%%%%%%%%%%%%%%%%%%%%%%%%
%%%%%%%%%%%%%%%%%%%%%%%%%%%%%%%%%%%%%%
\section{Final remarks}\label{sec:final}

In this work we analyzed the interplay between SC and NFL behavior at finite temperature, near a QCP for a Fermi surface coupled to a massless boson. We will end by summarizing the main points of our work, presenting the resulting phase diagrams, and discussing future directions.

We argued that the presence of infrared singularities from the exchange of zero-frequency bosons invalidates both the Eliashberg approximation and the standard replacement of the Matsubara sum by an integral (at low temperatures). This afflicts the fermionic self-energy and the SC gap. The resolution of IR divergences in the self-energy leads to a new thermal term $\Sigma_T$~\cite{Wang:2017kab}. Here we went beyond the Eliashberg approximation, deriving a gap equation that takes into account $\Sigma_T$ and that correctly includes the effect of thermal bosons with $\Omega_n=0$ (which is found to be subdominant at large $N$). The resolution of IR divergences turns out to have important physical implications. In particular, it leads to a parametric difference between the zero-temperature gap and $T_c$, $\Delta_0 \gg T_c$. It would then be interesting to understand if thermal fluctuations are responsible for a similar behavior observed in some strongly correlated materials.

The dynamics depends crucially on $N \epsilon$, with $N$ the rank of the $SU(N)$ global symmetry, and $\epsilon=3-d$. We obtained, both numerically and analytically, the SC critical temperature $T_c$ as a function of $N$. For $N \epsilon \ll 1$, NFL effects from the anomalous dimension are subleading, and $T_c$ is enhanced (compared to the BCS result) due to massless boson exchange. As $N \epsilon$ increases, NFL effects grow, and SC becomes more irrelevant. We found that at $N \epsilon =12$ the critical temperature vanishes with a BKT scaling behavior, $\ln{\left(T_c/\Lambda_\text{NFL}\right)} \sim -1/\sqrt{\frac{12}{\epsilon N}-1}$. For larger $N$, SC is extinguished in the critical regime and the system remains quantum critical down to zero temperature. SC only develops away from the critical regime, in which the inverse correlation length sets a ``Debye scale" for conventional BCS-like pairing.
 We summarize the resulting phase structure in Fig.~\ref{fig:phases}.

By increasing the number of flavors to $N \sim 12/\epsilon$ we obtain an NFL superconductor, where the quasiparticle description breaks down even above the SC dome. This provides a controlled theoretical framework for strange metallic behavior. The existence of naked quantum criticality for $N > 12/\epsilon$ is also quite tantalizing, especially for its thermodynamic and quantum entanglement properties. Our results, as well as previous works~\cite{Raghu:2015sna, Wang:2016hir}, suggest that proximity to a multi-critical BKT fixed point could play a role in strange metals.
It would be interesting to study further the phenomenology of this regime, for instance by looking at transport properties. We stress that our analysis hold at large $N$, but we hope that this can capture some of the physics for a small number of flavors, as expected in real materials. It will be interesting to quantify this further.

Our conclusions are consistent with the $T=0$ results in~\cite{Raghu:2015sna, Wang:2016hir}, and extend them to finite $T$. Indeed, it was found that the $T=0$ gap $\Delta_0$ is also strongly affected by NFL fluctuations as $N \sim N_c$, and that SC vanishes via an infinite order BKT-type quantum phase transition at $N=N_c$. It is noteworthy that this continuity of the phase structure down to $T=0$ can only be achieved by carefully including the effects of thermal fluctuations which, however, cannot be seen at the $T=0$ fixed point.

Finally, we discussed what happens if the strength of the NFL thermal term is artificially reduced. This allowed us to study the interplay between the special first Matsubara frequency and NFL effects. As a result, we recovered the Fermi liquid behavior obtained in~\cite{2016PhRvL.117o7001W}, as well as a new range with NFL-like first Matsubara physics. This suggests the intriguing possibility of re-entrant SC, where the gap is nonzero over a finite temperature interval, but quantum criticality emerges again at the lowest temperatures. We should stress that the artificial reduction of the self-energy cannot be achieved in the regime that was our focus here, which was based on an overdamped boson with $z=3$ dynamical exponent. Nevertheless, the interesting phenomenology that results motivates further explorations and, in particular, a discussion of SC at the nonlinear level. We hope to come back to these points in future work.

%%%%%%%%%%%%%%%%%%%%%%%%%%%%%%%%%%%%%%
%%%%%%%%%%%%%%%%%%%%%%%%%%%%%%%%%%%%%%
\acknowledgments{We thank Andrey V. Chubukov, Shamit Kachru, and Srinivas Raghu for stimulating discussions. 
H.W. is supported by DARPA YFA contract D15AP00108.
Y.W. is supported by the Gordon and Betty Moore Foundation's EPiQS Initiative through Grant No. GBMF4305 at the University of Illinois. 
G.T. is supported by CONICET, PIP grant 11220150100299, by ANPCYT PICT grant 2015-1224, and by CNEA.}

%%%%%%%%%%%%%%%%%%%%%%%%%%%%%%%%%%%%%%
%%%%%%%%%%%%%%%%%%%%%%%%%%%%%%%%%%%%%%
\appendix

\section{Derivation of the gap equation}\label{app:1}

In this Appendix we derive the gap equation (\ref{eq:gap}), which goes beyond the Eliashberg approximation. This is similar to the derivation of the fermion self-energy in \cite{Wang:2017kab}.

The quantity that we need to calculate in (\ref{eq:gap2}) is
\be
I_{mn} \equiv \int_0^\infty\,q^{1-\epsilon}dq\,D(\omega_m-\omega_n,q)\,\tan^{-1}\left(\frac{v q}{|Z(\omega_m)\omega_m|}\right)
\ee
and we will need the full boson propagator
\be
D( \Omega_n, q) = \frac{1}{\Omega_n^2+q^2+\frac{2 M_D^2}{\pi}\frac{\Omega_n}{q}\tan^{-1}\frac{vq}{\Omega_n}}\,.
\ee

Consider first $m \neq n$. We split the momentum integral into $vq< |\omega_m-\omega_n|$ and $vq >|\omega_m-\omega_n|$. In the first range, the boson propagator $D \sim 1/M_D^2$, and so the contribution from this regime is parametrically small in the low energy theory; we neglect it in what follows. Hence
\be
I_{mn} \approx \int_{|\omega_m-\omega_n|/v}^\infty dq \frac{q^{1-\epsilon} }{q^2+M_D^2  \frac{|\omega_m-\omega_n|}{q}}\,\tan^{-1}\frac{v q}{|Z(\omega_m)\omega_m|}\,.
\ee
Now we note that for small $\epsilon$, the inverse tan function here can be replaced by its limiting value $\pi/2$. Performing the $q$ integral then yields
\be
I_{mn}\approx  \frac{\pi}{2\epsilon}\,\frac{1 }{M_D^{2\epsilon/3}|\omega_m-\omega_n|^{\epsilon/3}}
\,.
\ee
On the other hand, the case $m=n$  may be evaluated explicitly to obtain
\be
I_{nn}=\frac{\pi}{2\epsilon}\,\frac{v^\epsilon}{|\omega_n+\Sigma(\omega_n)|^\epsilon}\,.
\ee
Replacing these results into (\ref{eq:gap2}) obtains (\ref{eq:gap}).

\section{Cancellation of diagonal term with $n=m$ in the gap equation}\label{app:2}

From (\ref{eq:sethermal}), (\ref{eq:gap}), the set of Eliashberg equations takes the form:
 \begin{align}
 \tilde\Delta(\omega_m)=&\frac{v^\epsilon\Lambda^{\epsilon/3}_{\rm NFL} M_D^{2\epsilon/3}\pi T}{|{\omega_m+\Sigma(\omega_m)|^{1+\epsilon}}}\frac{\tilde\Delta(\omega_m)}{N}\nonumber\\
 &+\frac{T}{N}\sum_{n\neq m}\frac{\tilde\Delta(\omega_n)}{|\omega_n+\Sigma(\omega_n)|} \frac{\Lambda_\text{NFL}^{\epsilon/3}}{|\omega_m-\omega_n|^{\epsilon/3}},\nonumber\\
 \Sigma(\omega_m)=&\frac{v^\epsilon\Lambda^{\epsilon/3}_{\rm NFL} M_D^{2\epsilon/3}\pi T}{|{\omega_m+\Sigma(\omega_m)|^\epsilon}}\sgn(\omega_m)\nonumber\\
 &+T\sum_{n\neq m}{\sgn(\omega_n)} \frac{\Lambda_\text{NFL}^{\epsilon/3}}{|\omega_m-\omega_n|^{\epsilon/3}}.
 \end{align}
We can rewrite the set of equations in a more compact form
 \begin{align}\label{eq: gap_compact}
 \tilde\Delta(\omega_m)=&\frac{T}{N}\sum_{n}\frac{\tilde\Delta(\omega_n)}{|\omega_n+\Sigma(\omega_n)|} u(\omega_m,\omega_n),\nonumber\\
 \Sigma(\omega_m)=&T\sum_{n}{\sgn(\omega_n)}u(\omega_m,\omega_n),
 \end{align}
 where 
 \be
 u(\omega_m,\omega_n)=u(\omega_m-\omega_n)=\frac{\Lambda_\text{NFL}^{\epsilon/3}}{|\omega_m-\omega_n|^{\epsilon/3}}
 \ee 
 for $m\neq n$, and 
 \be
 u(\omega_m,\omega_m)=\frac{v^\epsilon\Lambda^{\epsilon/3}_{\rm NFL} M_D^{2\epsilon/3}\pi}{|{\omega_m+\Sigma(\omega_m)|^\epsilon}}.\nonumber
 \ee 
 
In terms of the physical gap
 \be
 \Delta(\omega_m)\equiv \tilde\Delta(\omega_m)\omega_m/(\omega_m+\Sigma(\omega_m)),
 \ee
we have
 \begin{align}
  \Delta(\omega_m)\(1+\frac{\Sigma(\omega_m)}{\omega_m}\)=&\frac{T}{N}\sum_{n}\frac{\Delta(\omega_n)}{|\omega_n|} u(\omega_m,\omega_n).
  \label{11}
 \end{align}
 Let's substitute the expression for $\Sigma_m$ on the LHS, and single out the term $\omega_m=\omega_n$ on both sides of \eqref{11}: 
  \begin{align}
 & \Delta(\omega_m)\(1+\frac{\bar\Sigma(\omega_m)}{\omega_m}+T\frac{u(\omega_m,\omega_m)}{|\omega_m|}\)\nonumber\\
  &=\frac{T}{N}\sum_{n\neq m}\frac{\Delta(\omega_n)}{|\omega_n|} u(\omega_m,\omega_n)+\frac{T}{N}\frac{\Delta(\omega_m)}{|\omega_m|}u(\omega_m,\omega_m),
  \label{12}
 \end{align}
 where
 \begin{align}
  \bar\Sigma(\omega_m)=&T\sum_{n\neq m}{\sgn(\omega_n)}u(\omega_m,\omega_n),
\end{align}
 
 Reorganizing \eqref{12}, we get
 \begin{align}
& \bar \Delta(\omega_m)\[1+\frac{\bar\Sigma(\omega_m)}{\omega_m}+\(1-\frac{1}{N}\)T\frac{u(\omega_m,\omega_m)}{|\omega_m|}\]\nonumber\\
 &=\frac{T}{N}\sum_{\neq m}\frac{\Delta(\omega_n)}{|\omega_n|} u(\omega_m,\omega_n).
  \label{13}
\end{align}
We can now redefine through a ``reverse process"
\begin{align}
\tilde\Sigma(\omega_m)&\equiv\bar\Sigma(\omega_m)+(1-1/N)T\sgn(\omega_m) u(\omega_m,\omega_m)\nonumber\\
\tilde \Delta'(\omega_m)&\equiv \Delta(\omega_m)(\omega_m+\tilde\Sigma(\omega_m))/\omega_m,
\end{align}
 such that
 %\begin{widetext}
 \begin{align}
 \tilde\Delta'(\omega_m)=&\frac{T}{N}\sum_{n\neq m}\frac{\tilde\Delta'(\omega_n)}{|\omega_n+\tilde\Sigma(\omega_n)|} u(\omega_m,\omega_n),\nonumber\\
\tilde\Sigma(\omega_m)=&T\sum_{n\neq m}{\sgn(\omega_n)}u(\omega_m,\omega_n)\nonumber\\
&+\(1-\frac{1}{N}\)T\sgn(\omega_m) D(\omega_m,\omega_m).
\label{133}
 \end{align}
 %\end{widetext}
 What we have shown is that as an eigen-system, (\ref{eq: gap_compact}) is equivalent to (\ref{133}). The later eliminates the diagonal contribution to the gap equation, while modifying the coefficient ($1\to 1-1/N$) for the thermal contribution to the self-energy $\Sigma$. This justifies in large $N$ our treatment of the gap equation in the main section. 

\bibliography{NFL}
\bibliographystyle{utphys}

\end{document}